\documentclass{llncs}
\usepackage{graphicx}
\usepackage{amssymb}
\usepackage{amsmath}
\DeclareMathOperator{\expup}{e}
\newcommand{\etal}{et.~al.~}
\newcommand{\ket}[1]{\ensuremath{\left|#1\right\rangle}}

\newcommand{\Cre}[0]{\ensuremath{\hat{a}^{\dagger}}}
\newcommand{\Ann}[0]{\ensuremath{\hat{a}}}
\newcommand{\SP}[0]{\ensuremath{\hat{\sigma}^{+}}}
\newcommand{\SM}[0]{\ensuremath{\hat{\sigma}^{-}}}
\newcommand{\SThree}[0]{\ensuremath{\hat{\sigma}_3}}

\begin{document}
\title{Universal Continuous Variable Quantum Computation in the Micromaser}

\author{Rob C.~Wagner\inst{1}
\and Mark S.~Everitt\inst{2}
\and Viv M.~Kendon\inst{1}
\and Martin L.~Jones\inst{1}
}
\institute{School of Physics and Astronomy, University of Leeds, Leeds, UK, LS2 9JT
\and
National Institute of Informatics, 2-1-2 Hitotsubashi, Chiyoda-ku, Tokyo 101-8430, Japan
}
\maketitle
\begin{abstract}
We present universal continuous variable quantum computation (CVQC) in the micromaser. With a brief history as motivation we present the background theory and define universal CVQC. We then show how to generate a set of operations in the micromaser which can be used to achieve universal CVQC. It then follows that the micromaser is a potential architecture for CVQC but our proof is easily adaptable to other potential physical systems.
\end{abstract}

\section{Introduction}\label{intro}
Analogue computation has a long-running history, from the invention of the Astrolabe \cite{Morrison2007} for plotting the heavens in around 200 BC, through the slide rule and mechanical differential analyser 
to more modern electronic devices. Analogue computation is less well developed than its digital counterpart, but offers many opportunities both in the theoretical advancement and physical realisation of computers \cite{Shannon1941,Rubel1993,Moore1996}. 

Quantum mechanics is a more recent invention, being conceived and developed since the early 20th century. Much of the original research in quantum mechanics used continuous variable systems, such as operations on the positions and momenta of particles. It would thus appear that quantum mechanics offers a breeding ground for new theories of continuous variable (CV) computation. However, perhaps inspired by the prevalence of classical digital computation, most of the research into quantum computation is aimed at discrete variables, in the form of qubits \cite{Feynman1982,Deutsch1985,Feynman1986}. 

Lloyd and Braunstein \cite{Lloyd1999} laid the groundwork for continuous variable quantum computation (CVQC) in 1999. Almost immediately research went into implementing CV algorithms, such as analogues of Grover's, Deutsch \& Jozsa's and Shor's algorithms \cite{Pati2000,Pati2002,Lomonaco2002}, and investigating the general structure of computation and simulation with CVs. Some of these looked at implementing discrete computation embedded in CV systems \cite{Ralph2003,Spiller2006} and others looked at using the physics of CV systems directly to implement CV computing \cite{Bartlett2002,Kok2001}. 

There are two major schemes for CVQC: encoding the information in infinitely-squeezed states, such as the position eigenstates; or encoding the information in Gaussian states, such as the quantum coherent states of light. Previous work favours the first approach but we argue that the second is more appropriate for a realistic view and implementation of CVQC.

In this paper we will show how to achieve universal CVQC in a particular experiment, the micromaser. Computations and universality are described in Section \ref{UCVQC} with recipes for universality in two different encodings of variables. The micromaser itself is described in Section \ref{micromasers} along with the background physics. Our results are laid out in Section \ref{operations}, which is how to achieve universal CVQC in the micromaser and in Section \ref{summary} we give our plans for further work.

\section{Universal CVQC}\label{UCVQC}
The notion of universality is important in all branches of computation theory. We call our system a universal computer for our purposes if it can perform any computation for which we wish to use it. In discrete variable quantum computation (DVQC), universal computation is being able to achieve any unitary operation on the states encoding the variables. Since realising any unitary operation on a continuous variable would require an infinite number of parameters, for continuous variable quantum computation (CVQC) we restrict ourselves to exponentials of Hermitian polynomials on the space of continuous variables. This is sensible as unitary operations are usually considered to be the result of applying a Hamiltonian for a period of time and Hamiltonians are Hermitian polynomials 
\footnote{A polynomial $f$ in the position and momentum operators $\hat{x},\hat{p}$ is said to be Hermitian iff it is its own adjoint: $f^{\dagger}=f$}
. We encode the CV information in the eigenstates of some continuous-spectrum operator and computations are embodied as physical manipulations which correspond to operations on the eigenstates. This is the definition of continuous variable quantum computation that we employ. To encode our variables and to describe the physical modes involved we make use of the quadrature operators $\hat{x}$ and $\hat{p}$, which are orthogonal in the sense that $\left[\hat{x},\hat{p}\right]=i$, up to a real normalisation constant. Any polynomial in $\hat{x}$ and $\hat{p}$ can be generated given a certain set of available operators as stated by Lloyd \& Braunstein \cite{Lloyd1999}: \begin{quote} ``Simple linear operations on continuous variables, together with a non-linear operation and any interaction suffices to enact to an arbitrary degree of accuracy arbitrary Hermitian polynomials of the set of continuous variables.''\end{quote} 

In terms of operators, the simple linear operations are $\left\{ \pm\hat{x},\pm\hat{p}\right\} $, a non-linear operation might be the Kerr Hamiltonian $\hat{H}_{\mathrm{Kerr}}=\left(\hat{x}^{2}+\hat{p}^{2}\right)^{2}$ and an interaction is to couple modes together, e.g. two mode squeezing or the sum gate described later in this section. The way in which polynomials are generated is as follows: given Hamiltonians $\hat{A}$ and $\hat{B}$, for some small time $\delta t$
\begin{equation}
	\expup^{-i\hat{A}\delta t}\expup^{-i\hat{B}\delta t}\expup^{i\hat{A}\delta t}\expup^{i\hat{B}\delta t}=\expup^{\left[\hat{A},\hat{B}\right]\delta t^{2}}+\mathrm{O}\left(\delta t^{3}\right)
	\label{eq:1}
\end{equation}
and
\begin{equation}
	\expup^{i\hat{A}\frac{\delta t}{2}}\expup^{i\hat{B}\frac{\delta t}{2}}\expup^{i\hat{B}\frac{\delta t}{2}}\expup^{i\hat{A}\frac{\delta t}{2}}=\expup^{i\left(\hat{A}+\hat{B}\right)\delta t}+\mathrm{O}\left(\delta t^{3}\right)
	\label{eq:2}
\end{equation}
So we can generate the Hamiltonians $\pm i\left[\hat{A},\hat{\pm B}\right]$ and $\pm\hat{A}\pm\hat{B}$ from $\hat{A}$ and $\hat{B}$ very easily to arbitrary fidelity. Note that we are converting between the space of unitary operators and the space of Hamiltonians to generate our Hamiltonians since unitary evolution is how nature evolves. A non-linear operation $\hat{C}$ is one of order cubed or higher in $\hat{x}$ and $\hat{p}$, meaning that when it is commuted with another operator $\hat{H}$, an operator of higher order in $\hat{x}$ and $\hat{p}$ than $\hat{H}$ is obtained. 
Non-linear operations can't be efficiently simulated on a classical discrete variable computer \cite{Bartlett2002}.
Recursively then, polynomials of any order can be generated. 

While this gives us universality by our definition of obtaining any Hermitian polynomial in $\hat{x}$ and $\hat{p}$, there is some choice left in how to encode the CV quantities into the states. The set of continuous variable states is generally considered to belong to one of two distinct classes. One consists of variables encoded into a set of infinitely squeezed states - eigenstates of a quadrature operator - and the other is the set of Gaussian states. The infinitely squeezed states are a limit case of the Gaussian states but the two have slightly differing sets of universal operations. The two are not equivalent but lead to the same set of computable functions since we are looking for any Hermitian polynomial in the operators.

We choose to encode our information in finitely squeezed Gaussian states because infinitely squeezed states are somewhat unphysical. To achieve universal computation we need a non-linear operation (order cubed or higher) and the following list of linear operations (in order: displacement; fourier transform; 1-mode squeezing; 2-mode squeezing)
\begin{gather}
	\hat{X}\left(x\right)\equiv\exp\left(-2ix\hat{p}\right)\label{eq:7}\\
	\hat{F}\equiv\exp\left(\frac{i\pi}{2}\left(\hat{x}^{2}+\hat{p}^{2}\right)\right)\label{eq:8}\\	\hat{S}\left(\zeta\right)\equiv\exp\left(\frac{\zeta^{\star}}{2}\hat{a}^{2}-\frac{\zeta}{2}\hat{a}^{\dagger2}\right)\label{eq:9}\\	\hat{S}_{i,j}\left(\zeta\right)\equiv\exp\left(\zeta^{\star}\hat{a}_{i}\hat{a}_{j}-\zeta\hat{a}_{i}^{\dagger}\hat{a}_{j}^{\dagger}\right)\,,\label{eq:10}
\end{gather}
where $\Ann=(x+ip)/\sqrt{2}$ and $\Cre=(x-ip)/\sqrt{2}$ are the ladder operators. Here the first three are single-mode operations and the last is our interaction. We describe these operations in more detail in section \ref{operations}. For one-mode squeezing, Equation (\ref{eq:9}), the variable $\zeta\in\mathbb{C}$ describes the orientation and amount of squeezing (described later) and on the annihilation operator $\hat{a}$, $\hat{S}\left(\zeta\right)$ acts as 
\begin{equation}
	\hat{S}\left(\zeta\right)^{\dagger}\hat{a}\hat{S}\left(\zeta\right)=\expup^{+r}\hat{x}^{\left(\theta\right)}+i\expup^{-r}\hat{p}^{\left(\theta\right)},
\end{equation}
where
$\hat{x}^{\left(\theta\right)}=\hat{x}\cos\theta+\hat{p}\sin\theta$ and $\hat{p}^{\left(\theta\right)}=-\hat{x}\sin\theta+\hat{p}\cos\theta$
are the rotated quadratures. The last operation is the two-mode squeezing operator $\hat{S}_{i,j}\left(\zeta\right)$, which reduces the variance in the relative position and total momentum of the two modes being coupled.

\begin{figure}[ht]
	\centerline{
		\includegraphics{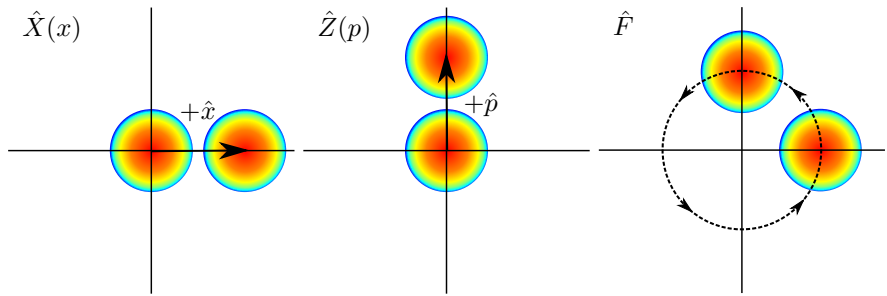}
	}
	\caption{A Q-function plot of positive displacements of position and momentum from the ground state to other coherent states. The third Figure demonstrates the Fourier transform for a rotation of $\pi/2$.}\label{fig2}
\end{figure}

\begin{figure}[ht]
	\centerline{
		\includegraphics{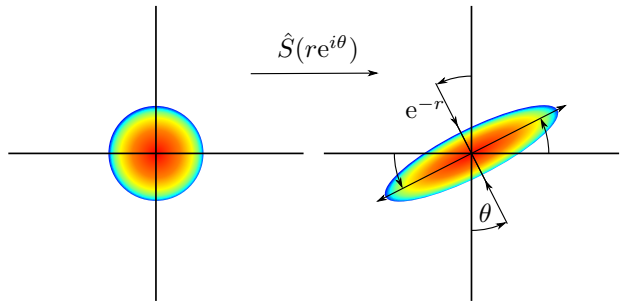}
	}
	\caption{One-mode squeezing applied to the ground state. Shown is how $r$ and $\theta$ parameterise the squeezing in phase space.}\label{fig4}
\end{figure}

One way to easily visualise the effect of these operations on the phase space of the modes is using Husimi Q-function plots. For example, in Figure \ref{fig2} we present the plots for the $\hat{X}\left(x\right)$ and $\hat{Z}\left(p\right)\equiv\exp\left(-2ix\hat{p}\right)$ displacement operations. The Fourier transform, $\hat{F}$ in Figure \ref{fig2}, gives a rotation of $\pi/2$ in phase space. We see that we can easily generalise the Fourier transform to $\hat{F}\left(t\right)\equiv\exp\left(it\left(\hat{x}^{2}+\hat{p}^{2}\right)\right)$. Figure \ref{fig4} shows the effect of one-mode squeezing applied to the ground state. The variance in one axis is increased while the variance in the conjugate axis is decreased to compensate. Two-mode squeezing generates a similar effect, but between the pair-wise positions and momenta of the two modes.
However, if we look at the relative positions and momenta instead, we see the correlations between the two modes. 

To achieve general universal CVQC, we need to go beyond the realms of Gaussian states (which are the eigenstates of the class of Hamiltonians only quadratic in $\hat{x}$ and $\hat{p}$). To do this requires a non-Gaussian operation, such as a non-linearity (see the above mentioned Kerr Hamiltonian), a cubic-phase gate \cite{Gottesman2001}, or simply measurement - although without a clever scheme this may not be very useful for computation due to the probabilistic nature of the process.

Having established a set of universal gates for computation, we briefly describe how to carry out a computation. We can create any eigenstate of an operator which can be generated by our universal set (equations \ref{eq:7}-\ref{eq:10} plus a non-linear operation). To do this we first initialise to an appropriate coherent state (say the ground state \ket{0}) and perform the appropriate sequence of operations which generates the desired unitary operation. We have shown (equations \ref{eq:1} \& \ref{eq:2}) that we can generate any desired unitary operation from the space of those available to us starting with our basis set but in practice there could be a long sequence of elementary operations to achieve the desired operation.
After applying the desired operation we can use homodyne-like measurements to find the output state and hence determine the result of the computation. The feasibility of an individual computation is at the mercy of the comparison of the resources available and the resources required, such as coherence times, number of modes etc.

\section{What is a micromaser?}\label{micromasers}
\begin{figure}[ht]
	\centerline{
		\includegraphics{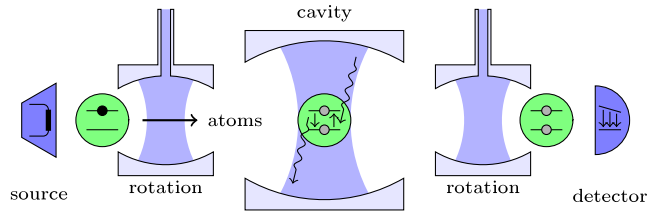}
	}
	\caption[Conceptual phase sensitive micromaser.]{The `phase sensitive' micromaser. Atoms are emitted by the source in the excited state and may be rotated coherently between the ground and excited states (both of which are actually highly excited Rydberg states of the atom) with a lossy microwave field, labelled as rotation here. The cavity in the centre may then interact with the atom, and a final rotation on the atom allows a measurement basis to be selected for measurement of the atomic state using state selective field ionisation.\label{PhaseSensitiveConcept}}
\end{figure}

To perform CV quantum computations we propose the use of a micromaser-like system. The micromaser \cite{Meschede1985}, or microscopically pumped maser, is essentially a very high quality microwave cavity with a rarified beam of atoms passing through it. The transition of single atoms moving through the cavity one at a time from one highly excited \emph{Rydberg} state to another will keep the microwave field in the cavity pumped. This is an extreme case of a beam maser, which would normally use a dense beam of atoms to pump a lossier cavity. The micromaser has been historically important as a test of cavity QED as a physical realisation of the Jaynes-Cummings model \cite{Jaynes1963} which predicts the behaviour of a two level atom in a single mode field without semi-classical approximations. This model is important as the foundation of our understanding of masers and lasers. A simplified schematic can be found in Figure \ref{PhaseSensitiveConcept}.

By virtue of the very high quality of a micromaser cavity, for which the best examples may retain a photon for 0.3\,s  \cite{Walther2006}, the linewidth of the cavity mode is very small. For cavity of lifetime 0.3\,s the linewidth is less than a hertz for a mode frequency of 21.456\,GHz. The field in a micromaser cavity is well suited to CVQC due to this long lifetime and well defined mode frequency. Quantum coherence of the cavity field is also maintained by the high quality cavity. We propose the use of the quadratures of the cavity field as a continuous variable for CVQC. Multi-variable computations require extensions of the micromaser to multi-mode fields which requires modified cavity designs and carefully chosen energy levels of the atom. The Rydberg atoms that are commonly used in micromaser experiments provide a multitude of possible transitions to couple to many modes.

\subsection{The Jaynes-Cummings Model}
The Jaynes-Cummings model describes the simplest non-trivial interaction between atoms and light, consisting of a single atom with two states, labelled \ket{e} for the upper and \ket{g} for the lower\footnote{These stand for \emph{excited} and \emph{ground} respectively, although in a laboratory two highly excited Rydberg states of rubidium are chosen such that the difference in energy between the states is close to resonance with the microwave cavity field.}, interacting with a single mode field. The Hamiltonian of this interaction is \cite{BarnettRadmore}
\begin{equation}\label{eq:17}
\hat{H}=
\underbrace{\frac{\hbar\omega_{\mathrm{a}}}{2}\SThree		\rule[-8pt]{0pt}{5pt}}_{\mbox{atom}}
+\underbrace{\hbar\omega\Cre\Ann						\rule[-8pt]{0pt}{5pt}}_{\mbox{field}}
-\underbrace{i\hbar g\left(\SP\Ann-\SM\Cre\right)			\rule[-8pt]{0pt}{5pt}}_{\mbox{interaction}}\,,
\end{equation}
where $\hbar\omega_{a}$ is the transition energy of the atom from \ket{g} to \ket{e}, $\omega$ is the frequency of the mode, $g$ is the atom-field coupling constant, \Cre (\Ann) is the creation (annihilation) operator of the field mode and \SP (\SM) is the atomic raising (lowering) operator. This model can be generalised easily to include more field modes or atomic levels.

\subsection{Some experimental considerations}
The source of atoms used in a micromaser is typically an effusive oven. The atoms emitted will thus have Poisson distributed arrival statistics and a velocity distribution determined by the oven temperature. The velocity distribution can be refined by the use of Fizeau wheel or detuning the first step excitation laser and placing it at an angle to the atomic beam \cite{Englert2002} to select the velocity which Doppler shifts the laser into resonance. Poisson statistics of the atomic beam mean that there is some chance of multiple atom events in the cavity which are undesirable. The strategy used is to rarify the beam such that the chance of two or more atom events is lowered. The probability of a one-atom (desirable) event is given by $P_1=\expup^{-2rL/v}$ \cite{Englert2002},
where $r$ is the rate of atomic arrivals at the cavity, $L$ is the length of the cavity and $v$ is the velocity of the atoms. Given that a typical micromaser cavity is approximately 3\,cm long and that the selected velocity of the beam will be approximately 300\,m\,s${}^{-1}$, an average rate of 10 atoms per second will result in the probability of an event being a one atom event of $99.8\%$. Operations that require many atoms may have to sacrifice some precision due to the the greater atomic rates necessary to complete a computation within the decoherence time of the cavity field. For experiments run by the micromaser group in Garching, cavities as good as up to a quality factor of $Q=4\times10^{10}$ were used, corresponding to a life time of the field of 0.3\,s.

Some of the most recent advances in micromaser physics have involved the generation of Fock (number) states \cite{Brattke2001}. Using a micromaser field in a Fock state it is possible to produce a single atom in a particular state on demand \cite{Walther2006}. If atoms on demand can be supplied then the rate can be greatly increased as multiple atom events are effectively eliminated, and many more computational operations performed. Another intriguing possibility for a single atom source is with the use of a standing-wave dipole trap, which can be used to accelerate single atoms deterministically \cite{Kuhr2001}.

\section{Operations in the Micromaser}\label{operations}
Having described the two main schemes for encoding CVs in quantum systems for computation in Section \ref{UCVQC}, we need to choose a scheme for use in the micromaser. To decide this, it is necessary to look first at which states we can initialise and what operations we can perform on them. Coherent states\footnote{We use the standard definition of a coherent state, $\ket{\alpha}=\expup^{-\left|\alpha\right|^{2}/2}\sum_{n}\frac{\left(\alpha\right)^{n}}{\sqrt{n!}}\ket{n}$.} \cite{BarnettRadmore}, number (Fock) states (including Trapping States) \cite{Brattke2001}, steady states \cite{Weidinger1999}, and tangent \& cotangent states \cite{Slosser1990} may all be produced in a micromaser cavity field.
Most of these are unsuitable for UCVQC but we see that we can create coherent states. We now show how we can achieve UCVQC on the Gaussian states, going through the list of required operations.

\subsection{Displacement Operations}
The single-quadrature displacement functions
\begin{equation}
	\hat{X}\left(x\right)\equiv\expup^{-2ix\hat{p}}\,,\ \hat{Z}\left(p\right)\equiv\expup^{2ip\hat{x}}
	\label{eq:23}
\end{equation}
are of the form of the generalised Displacement Operator,
\begin{equation}
	\hat{D}\left(\alpha\equiv x+ip\right)\equiv\exp\left(2ip\hat{x}-2ix\hat{p}\right).
	\label{eq:25}
\end{equation}
Explicitly, $\hat{X}\left(x\right)=\hat{D}\left(x\right)$ and $\hat{Z}\left(p\right)=\hat{D}\left(ip\right)$. These operations are depicted in Figure \ref{fig2}.

Displacement is remarkably simple to achieve in the micromaser cavity field \cite{Agarwal1993,Lange1993}. By applying the appropriate external coherent field we can displace the state in the cavity by $\hat{D}\left(\alpha\right)$, thus by choosing an appropriate $\alpha$, we can achieve $\hat{X}\left(x\right)$ and $\hat{Z}\left(p\right)$ in the micromaser. The experiments of Lange and Walther \cite{Lange1993} demonstrated control over the average photon number over a large domain, from the subphoton level for states less than $\alpha=100$. Via feedback we can make these as stable as we require \cite{Engen1958}. With modern microwave synthesisers much better performance is expected, however without knowledge of the device used it is difficult to estimate this improvement over the performance in the original experiment. In an actual experiment the power required for specific shifts should be determined experimentally as each cavity will couple at a different strength to an external field due to differences in machining.

\subsection{The Fourier Transform}
The Fourier transform $\hat{F}\equiv\expup^{i\pi\left(\hat{x}^{2}+\hat{p}^{2}\right)/2}=\expup^{i\pi\left(\hat{N}+1\right)/2}$ 
is simply a $\pi/2$ rotation in phase space about the origin. For example, on a ground state displaced in the positive-$\hat{x}$ direction, we obtain a ground state displaced in the positive-$\hat{p}$ direction, as shown in Figure \ref{fig2} in Section \ref{UCVQC}. It has a trivial class of eigenstates, the number states:
\begin{equation}
	\hat{F}\ket{n}=\expup^{i\pi\left(n+1\right)/2}\ket{n}=i^{n+1}\ket{n}\,.
	\label{eq:26}
\end{equation}

This can easily be achieved using a micromaser with the atom detuned from the mode which we wish to apply the Fourier transform to, and far detuned from the other modes so as not to act on those. Given that the linewidth of each mode is so relatively small, this is a trivial requirement. For a mode detuned by $\Delta$ from an atomic transition the atom does not make a transition from \ket{e} to \ket{g}, but the combined system evolves for a time $t$ as
\begin{equation}
\ket{e,n}\mapsto\expup^{-ig^2(n+1)t/\Delta}\ket{e,n}\,,
\end{equation}
and the atom may be neglected after interaction. The detuning can clearly be chosen to satisfy the Fourier transform in Equation \eqref{eq:26}. 

\subsection{One-Mode Squeezing}
As there is no direct squeezing operation in the micromaser, we will first describe squeezing for linear optics since the formalism is the same.
In linear optics, one-mode squeezing
\begin{equation}
	\hat{S}\left(\zeta\right)\equiv\exp\left(\frac{\zeta^{\star}}{2}\hat{a}^{2}-\frac{\zeta}{2}\hat{a}^{\dagger2}\right)
	\label{eq:29}
\end{equation}
is generated via a nonlinear-optical $\chi^{\left(2\right)}$ interaction and yields the attenuation of a quadrature and the amplification of its conjugate. This is called nonlinear in optics, but to us this is a linear operation since it's only quadratic in the field operators. We'll still need a higher-order nonlinearity for our universal gate set. 

Since we are not considering linear optics, an analogue must be found for a micromaser system. The simplest system to consider is the two mode micromaser. This is best described as a three level ladder of atomic states for which the difference in energy between the uppermost and the lowermost states is twice the frequency of the field. The middle state is detuned from the one photon transition. 
As the central state is detuned from resonance with the field, the effective process is two photon transitions between the upper and lower state of the atom. The atom must be prepared in a particular superposition and after interaction the atom must be pulsed with a classical field so that it does not `give away' information about the cavity field and lead to decoherence. 

The particulars of this system were discussed in a paper by Orszag \etal \cite{Orszag1992}, but it is intuitive that $\hat{H}_i=\kappa\left(\hat{a}^2-\hat{a}^{\dagger2}\right)$ will be the form of the two photon process. By analysing the cavity's steady state, if it begins in a superposition ${\sum_{n}}A_{n}|n\rangle$, then it tends towards a squeezed state if $\left|\alpha/\gamma\right|=1-\epsilon$ for a very small $\epsilon$. This state is highly squeezed in momentum but is not quite of minimum uncertainty. We cannot take an infinite limit as the amount of squeezing is dependent on $\epsilon$ and all the squeezing disappears for $\epsilon=0$ but we can get an arbitrary amount of squeezing. After many atoms have passed through the cavity, we approach a pure squeezed state with $\left(\Delta\hat{p}\right)^{2}\simeq0$.

Since it can take a long time to achieve this amount of squeezing, we may wish to prepare a sufficient quantity of squeezed states in advance in order that we may transfer the squeezing onto a state during the computation. Lam \etal \cite{Lam1999} have demonstrated this using Optical Parametric Oscillators in linear optics with the crucial component of a beam splitter to couple the modes. We show in Section \ref{2mode} that a beam splitter may be replaced by any interaction and then explain an achievable interaction for the micromaser.

\subsection{Two-Mode Squeezing}\label{2mode}
In linear optics, two-mode Squeezing
\begin{equation}
	\hat{S}_{i,j}\left(\zeta\right)\equiv\exp\left(\zeta^{\star}\hat{a}_{i}\hat{a}_{j}-\zeta\hat{a}_{i}^{\dagger}\hat{a}_{j}^{\dagger}\right)
	\label{eq:34}
\end{equation}
is generated via a non-degenerate optical parametric amplifier (NOPA). It generates correlations between two modes, so it is an entangling operation. 

\begin{figure}[ht]
\centerline{
\includegraphics{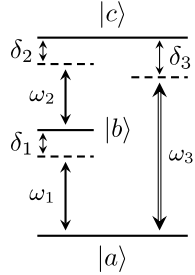}
}
\caption{The energy level diagram for the 3-level atom. $\delta_n$ is the detuning of the relevant mode from each transition, and $\omega_n$ is the frequency of each mode. The thick arrow between \ket{a} and \ket{c} denotes a classical pump field. If the atom moves up the ladder from $\ket{a}\mapsto\ket{b}\mapsto\ket{c}$ it will then be pumped back down to \ket{a} by the classical field, acting on the field as the net operator $\Ann_1\Ann_2$. The converse is also true so that this system will behave like two-mode squeezing.\label{ThreeLevelSqueezing}}
\end{figure}

While it is a simple operation in mathematical terms, it is not as easy to achieve in the micromaser. However, the result for universality given in Section \ref{UCVQC} tells us that we need \emph{any interaction} between two modes. By passing a 3-level atom with the energy level diagram in Figure \ref{ThreeLevelSqueezing} through a cavity with two modes we can get an interaction between the modes which effectively excludes the atom.
The interaction also involves a coherently pumped lossy field and allows photon transfer between the two relevant modes through virtual processes. It is governed by the effective Hamiltonian with the assumption that $\delta_1,\delta_2,\delta_3\gg g_1,g_2,\Gamma$
\begin{equation}
\hat{H}_{\mathrm{eff}}=\Theta+\frac{ig_1g_2\Gamma}{\hat{P}\hat{Q}}\left(\Cre_1\Cre_2-\Ann_1\Ann_2\right)\,,
\end{equation}
where 
$\Theta$ acts as a Fourier transform component on each mode. The rest is clearly related to two mode squeezing as the Schr\"odinger equation is solved to give the evolution operator $\hat{U}=\expup^{-i\hat{H}_{\mathrm{eff}}t}$. The coupling strength $g_1$ ($g_2$) is between the first (second) cavity mode and the transition between states \ket{a} and \ket{b} (\ket{b} and \ket{c}) of the atom, $\Gamma$ is the coupling strength between the coherently pumped lossy field and the transition between levels \ket{a} and \ket{c} and
\begin{equation}
\hat{P}=\sum_{i=1,2}\Cre_i\Ann_i\left(\frac{\delta_3}{2}-\delta_i\right),\ \hat{Q}=\hat{P}+\delta_2\,.
\end{equation}
In an actual experiment this final transition would likely be replaced with a two photon transition to follow selection rules.
We need to modify the state vector simultaneously by $\ket{\psi^\prime}=\exp\left(i\sum_{i=1,2}\Cre_i\Ann_i\left(\omega_{i}+\delta_{i}-\frac{\delta_{3}}{2}\right)t\right)\ket{\psi}$.
Now, two modes inside a single cavity can be coupled and we have an interaction polynomial which is good enough.. The effective coupling constant of the interaction is $g_1g_2\Gamma/\hat{P}\hat{Q}$ and the interaction is very close to two mode squeezing.

\subsection{Non-linearity}
As well as the simple linear and interaction terms listed above, we need some non-linearity to achieve universal (not just Gaussian state) computation. We can simply measure the state of a 2-level atom (see Section \ref{micromasers}) after transit through the cavity. This does not give a clean Hamiltonian, but any non-linearity is needed to be able to generate any Hermitian polynomial of continuous variables.

\section{Summary and Future Work}\label{summary}
In this paper we have given an overview of the history of continuous variables in computation, both classical and quantum. We stated the standard result for universal continuous variable quantum computation (CVQC) for two different encodings. We then gave an account of the micromaser, both what we can do in the experiment and what states can be created in the cavity. Given this, we showed how we can in principle achieve universal CVQC using Gaussian states in the micromaser using simple interactions.

Having a system which can, in principle, perform universal CVQC is very useful but we must consider how feasible it is to use the micromaser in such a way. All the necessary operations can be produced quickly and accurately in the system and any Hamiltonian can be generated by a polynomial number of the base operations. We may have a problem of scalability since our current interaction is between two modes in one cavity. However, for low numbers of modes the micromaser is a perfect candidate for efficient universal CVQC.

\section{Acknowledgments}
We thank Bill Munro and Kae Nemoto for first suggesting that CVQC was significant and interesting. We thank our funders: RCW is funded by the UK EPSRC; MSE is funded by the Japanese Society for the Promotion of Science; VMK is funded by a UK Royal Society University Research Fellowship.

\end{document}